\documentstyle[12pt,aps,amsmath,epsfig]{revtex}
\tightenlines
\begin{document}

\title{Nonlinear Quantum Transport and Current Noise}
\author{
  Gunther Lang  and Ulrich Weiss  }
 \address{ 
      II. Institut f\"{u}r Theoretische Physik, Universit\"{a}t
           Stuttgart, 70550 Stuttgart, Germany\\
 }
\address{\rm email: weiss@theo2.physik.uni-stuttgart.de}
\maketitle

\begin{abstract}
We study nonequilibrium transport and noise in a generic dissipative
tight-binding model. 
Within a real-time path integral
approach, we derive formally exact series expressions  in the number
of tunneling events for the noise valid for arbitrary bias,
frequency and temperature. At zero temperature, the low-frequency noise
can be summed in analytic form. The resulting Shiba-like
$|\omega|$-singularity  is a consequence 
of the $1/t^2$ decay law for the current correlation function. At
finite temperature, this singularity is smoothed out.
\end{abstract}

\pacs{PACS numbers: 05.60.Gg, 71.10.Pm, 73.40.Gk }
\narrowtext

Low-energy excitations of 1D conduction electrons and of edge currents
in fractional quantum Hall (FQH) systems \cite{wen}
are bosonic and constitute prime 
realizations of a Luttinger liquid (LL) phase of interacting fermions
\cite{haldane,schulz}.
A sensitive probe of this collective state is the tunneling conductance
of electrons through an impurity or point contact in a 1D quantum wire
\cite{kafi}, and of
Laughlin quasiparticles through a narrow constriction for
FQH edge currents \cite{fendley1}. 
In these systems, recent attention has been paid to
current fluctuations with emphasis on dc noise. To second order in the
tunneling amplitude, the dc noise $S$ at $T=0$ is classical shot noise, 
$S=2e^*I$ \cite{kane94,chamon1}. Thus upon measuring dc current and dc noise
in the weak-tunneling limit, one can determine the charge $e^*$ of the 
tunneling entity. In this way, the fractional charge of the Laughlin 
quasiparticles has been manifested \cite{glattli,weizmann}.
In the strong tunneling regime, higher order tunneling processes
cause a deviation of the dc noise from shot noise 
\cite{fendley2,solid}. Furthermore, correlations between  
tunneling events lead to colored noise.
Singular behavior  $S(\omega)\propto |\omega|$ has been found 
within a nonperturbative analysis of the equivalent 
integrable Boundary Sine-Gordon (BSG)
model \cite{saleur,chamon3}.
In this treatise, the prefactor of the $|\omega|$-law depends on the
particular treatment of the reflection matrix for quasiparticle
scattering at the boundary coming from the Fermi sea (cf. the debate
in Refs. \cite{saleur,chamon3}). 
 
In this paper, we study the low-frequency noise  upon employing 
real-time path integral techniques to a generic 
quantum transport model introduced by Schmid \cite{schmid},
\begin{align}
\label{HTB}
H &=H_{\rm TB}
+ \sum_{\alpha} \bigg[\,\frac{p_\alpha^2}{2m_\alpha} + \frac{1}{2} 
        m_\alpha \omega_\alpha^2 \left(x_\alpha 
        - \frac{c_\alpha}{m_\alpha \omega_\alpha^2} q\right)^2
\bigg] \;.
\end{align}
The Hamiltonian (\ref{HTB}) describes a particle which moves in the 
tilted tight-binding (TB) lattice
$H_{\rm TB}= -\frac{1}{2}\hbar\Delta\sum_{n}(c_n^\dagger c_{n+1}
+c_{n+1}^\dagger c_n) -\hbar \epsilon q/q_0$
with lattice spacing $q_0$, tunneling matrix element $\Delta$,
and biasing force $F=\hbar\epsilon/q_0$. 
The particle is bilinearly coupled via its 
position $q=q_0\sum_n c_n^\dagger c_n$ to a bath of
delocalized harmonic bosons. In the reduced density matrix description for
the Brownian quantum particle,
all effects of the environmental coupling are captured by the spectral density 
$J(\omega) = (q_0^2/2\hbar)\sum_\alpha (c_\alpha^2/m_\alpha\omega_\alpha^2)
\delta(\omega-\omega_\alpha)$. 
The TB model (\ref{HTB}) is dual to a corresponding weak-binding (WB)
model, in which the particle moves in a tilted cosine potential.
In the duality, the power $s$ in the generic form $J(\omega)\propto \omega^s$ 
is transformed into $2-s$.
Thus, super-Ohmic friction ($2>s>1$) is mapped on sub-Ohmic
friction ($0<s<1$), and vice versa \cite{wbuch}.
In the Ohmic case, $J(\omega) = 2K\omega$, the TB and WB models are
selfdual to each other. The dimensionless Kondo parameter $K$ is mapped
onto the reciprocal value $1/K$, and the bias $\epsilon$ onto
$\epsilon/K$,  as first observed by Schmid \cite{schmid}.

The TB and WB model describe many transport problems in solid state physics. 
Regarding applications of the model (\ref{HTB}) to the beforementioned
case of tunneling of a charge  in a LL phase, the spectral 
density is strictly Ohmic for $\omega \lesssim \omega_c$.
In the {\it scaling limit},
$\omega_c$ is by far the largest frequency of the problem.
The interaction parameter $g$ of the Luttinger model corresponds to
$1/K$, and the bias in the TB model (\ref{HTB}) is related to the voltage 
drop $V$
across the weak link or constriction by $\hbar\epsilon=eV$, where $e$ is 
the unit charge.
Another important example where the model (\ref{HTB}) applies
is a voltage-biased Josephson junction \cite{schoen,ing}. 
Here, the WB and TB limits 
correspond to the charge and phase representation, respectively.

Fluctuations are measured relative  to a stationary
state. Therefore, we proceed as follows. We first presuppose that the
global system is in a factorized system-bath state in the infinite past.
Then we switch on the bath coupling and the tilting force,  and let the 
system evolve under reign of the full Hamiltonian (\ref{HTB}). 
By the time $t=0$, the  system has reached a stationary nonequilibrium state, 
indicated by the index ``ne''.
The fluctuations of the current $j(t)=e \dot q(t)/q_0$ are reflected in
the noise spectrum, which is the Fourier transform of the symmetrized
current correlation function,
\begin{equation}\label{defnoise}
S(\omega)=\int_{-\infty}^{\infty}d\omega\,e^{i\omega t} \,\langle
j(t)j(0)+j(0)j(t)\rangle_{\rm ne} \;.
\end{equation}
Observing that the current 
correlation function is related to the mean square displacement $C_{\rm
ne}(t)=\langle[q(t)-q(0)]^2\rangle_{\rm ne}$ by
\begin{align}\label{current1}
\langle j(t)j(0)+j(0)j(t)\rangle_{\rm
ne}= \frac{e^2}{q_0^2}\,\frac{\partial^2}{\partial t^2}\, C_{\rm ne}(t) \;,
\end{align}
we can express the noise $S(\omega)$ in terms of the spectral function 
of $C_{\rm ne}(t)$ as
\begin{align}
  \label{sauscne}
S(\omega)&=-\frac{e^2}{q_0^2}\omega^2\widetilde{C}_{\rm ne}(\omega) \;,&
\widetilde{C}_{\rm ne}(\omega)&=\int_{-\infty}^{\infty}\!dt\,
e^{i\omega t}C_{\rm ne}(t)\;.   
\end{align} 
In particular parameter regimes, the noise can directly be expressed in
terms of transport quantities, e.\,g. the shot noise $S=2eVG(V)$ and
Johnson-Nyquist noise $S=4k_{\rm B}TG(V=0)$, where $G(V)=I(V)/V$ is
the nonlinear conductance. For later convenience, we also introduce
the differential conductance $G_{\rm d}(V)=\partial I(V)/\partial V$. The
conductances are related to the mobilities of the model (\ref{HTB}), 
$\mu(F)=\langle \dot q(t\to \infty)\rangle_{\rm fc}/F$ and  $\mu_{\rm
d}(F)=\partial \langle \dot q(t\to\infty)\rangle_{\rm fc}/\partial F$, by
$G=e^2\mu/q_0^2$ and $G_{\rm d}=e^2\mu_{\rm d}/q_0^2$. 
The index ``fc'' indicates that we may choose for the system-bath
state at initial time zero a factorized form. This is possible since
the leading contribution to mean values at asymptotic times
$t\to\infty$ does not depend on system-bath correlations in the
initial state. 

Within nonequilibrium linear 
response theory, the nonequilibrium susceptibility
\begin{align}\label{defchine}
\chi_{\rm ne}(t)&=\frac{i}{\hbar}\Theta(t)\,\langle
q(t)q(0)-q(0)q(t)\rangle_{\rm ne} \;, & \tilde{\chi}_{\rm
ne}(\omega)&=\int_{-\infty}^\infty dt\,e^{i\omega t}\chi_{\rm ne}(t)
\end{align}
comprises the differential mobility as 
\begin{align}\label{diffmobauschi}
\mu_{\rm d}(F)&=\lim_{\omega\to
0}\big[-i\omega\tilde\chi_{\rm ne}(\omega)\big]
 =\lim_{t\to\infty}\chi_{\rm ne}(t)\;.
\end{align}

Formally exact series expressions in $\Delta$ for the above quantities
can be derived 
upon employing a real-time path integral method for the reduced
density matrix (RDM). Because of the discreteness
of the TB model, a simple picture of a grandcanonical gas of
charges can be developed. The charges are stringed at the flip times
$\{t_j\}$, and the effects of the environment are in the charge
interactions. For a path with $n$ flips in
the negative-time branch ($t_1,\,t_2,\cdots,t_n<0$)
and $m$ flips in the positive-time branch 
($t_{n+1},\, t_{n+2},\dots, t_{n+m} >0$)
labelled by
charges $\{\xi_j=\pm 1\},\;\{\eta_j=\pm 1\}$, the interaction factor or
influence function is $F_{n,m}=G_{n,m}H_{n,m}$, where
\begin{align}
  \label{Gij}
  G_{n,m}&=\exp\bigg\{\! \sum_{k=2}^{n+m} \sum_{j=1}^{k-1}
\xi_k\xi_j\,Q'(t_k-t_j)\! \bigg\}\;, &
H_{n,m}&=\exp\bigg\{i\!\sum_{j=1}^{n+m-1} \eta_j \theta_{j,n+m}\bigg\}
\end{align} 
with $\theta_{j,n+m}=\sum_{k=j+1}^{n+m}\xi_k Q''(t_k-t_j)$. The
charges $\{\eta_j\}$ describe propagation on the infinite RDM in
diagonal direction, whereas the charges $\{\xi_j\}$ indicate moves
perpendicular to the diagonal, i.\,e. represent quantum
fluctuations. The pair interaction $Q(t)$ is the second integral of
the autocorrelation function of the random force $f(t)=\sum_\alpha
c_\alpha x_\alpha(t)$ in thermal equilibrium, $\ddot
Q(t)=(q_0/\hbar)^2 \langle f(t)f(0)\rangle_{\rm eq}$ \cite{wbuch}. 
In the Ohmic scaling limit, the pair interaction reads 
\begin{align}
\label{Qscal}
Q(t) &= Q'(t)+iQ''(t)= 2K\,\ln \big[(\hbar \beta \omega_c/\pi)
      \sinh(\pi |t|/\hbar \beta) \big]
        +i\pi K\,{\rm sgn}(t)\;.
\end{align} 
The influence of the bias $\epsilon$ or voltage drop $V$
is included in the bias phase factor
\begin{align}
  \label{biasfaktor}
 b_{n,m}&=\exp(-i\phi_{n,m})\;,& \phi_{n,m}&=\epsilon\sum_{l=1}^{n+m}
\xi_l t_l =\epsilon \sum_{l=1}^{n+m-1}g_{l,n+m}\,\rho_l \;. 
\end{align}
Here, we have introduced as an off-diagonal measure
the cumulative charge after $l$ jumps
$g_{l,n+m}=\sum_{k=l+1}^{n+m}\xi_k=-\sum_{k=1}^{l}\xi_k$, and 
the time intervals $\rho_l=t_{l+1}-t_l$. 

Following the lines of Ref.~\cite{wbuch}, it is straightforward
to  arrive at formally exact series expressions for the Laplace transforms of 
the above quantities in which the summation over all possible sequences 
$\{\eta_j\}$ has been performed already.

The first moment in the stationary nonequilibrium state 
with initial condition $\langle q(t=0)\rangle_{\rm ne} =0$ is given by 
\begin{align} 
\label{qnelambda}
\langle\hat q(\lambda)\rangle_{\text {ne}}&=\langle\hat{q}(\lambda)
\rangle_{\text {fc}}+\hat{R}_{\text {ne}}(\lambda)\;.
\end{align}
The term $\langle\hat{q}(\lambda)\rangle_{\text {fc}}$ is the first
moment with factorized system-bath initial state at time zero. This
term has dynamics  only in the positive-time branch.  
The residual term $\hat{R}_{\text {ne}}(\lambda)$ describes the corrections
due to the correlated initial state,
\begin{align}
\label{qlambdafc}
\langle\hat{q}(\lambda)\rangle_{\text
{fc}}&=\frac{-i}{\lambda^2}\sum_{m\; \rm even}^\infty
\int_{0}^\infty\!\tilde{\mathcal{D}}_{0,m}(\lambda)
\sum_{\{\xi_l\}}a_{0,m}^{(1)} G_{0,m} \sin\phi_{0,m} \;,\\
\label{rnelambda}
\hat{R}_{\text{ne}}(\lambda)&=\frac{-i}{\lambda} \sum_{n,m=1}^\infty
\int_{0}^\infty\!\tilde{\mathcal{D}}_{n,m}(\lambda)\int_0^\infty
\!d\tau_n\,ds_n \,e^{-\lambda s_n}
\sum_{\{\xi_l\}}a_{n,m}^{(1)} G_{n,m} \sin\phi_{n,m}\;.
\end{align}
The sum is over all charge sequences $\{\xi_l\}$ which form a
neutral cluster, $\sum_{l=1}^{n+m}\xi_l=0$.
The integration symbol includes the $\Delta$-factors associated with
each jump and the integrations over the interval lengths except for
the interval $\rho_n$, 
\begin{equation}
   \int_{0}^\infty\!\tilde{\mathcal{D}}_{n,m}(\lambda)\;\cdots=\Delta^{n+m}
\bigg(\prod_{l=n+1}^{n+m-1} \int_0^\infty d\rho_l\, e^{-\lambda \rho_l}\bigg)
\prod_{k=1}^{n-1} \int_0^\infty d\rho_k \;\cdots\;. 
\end{equation}
 The interval $\rho_n$ around $t=0$ is divided into the
negative-time part $\tau_n=-t_n$ and the positive-time part
$s_n=t_{n+1}$ and is treated separately. Finally, the 
phase factor  
\begin{align}
 a_{n,m}^{(1)}&=i\frac{q_0}{2}(-1)^{(n+m-2)/2}\prod_{j=1}^{n+m-1}
\sin(\theta_{j,n+m})
\end{align}
subsumes the interaction between the $\{\eta_j\}$ and
$\{\xi_j\}$ charges. In the Ohmic scaling limit, the phase
$\theta_{j,n+m}$ is independent of the flip times and is  
$\theta_{j,n+m}=\pi K g_{j,n+m}$. 

Similarly, the series expression for the nonequilibrium susceptibility reads
\begin{align}\nonumber 
 \hat{\chi}_{\text {ne}}(\lambda)&=\frac{q_0}{i\hbar\lambda} \!
\sum_{n,m=1}^\infty \!
\int_{0}^\infty\!\!\!\tilde{\mathcal{D}}_{n,m}(\lambda)\!\int_0^\infty
\!\!\!d\tau_n\,ds_n \,e^{-\lambda s_n}\!\sum_{\{\xi_l\}} g_{n,n+m}\,
a_{n,m}^{(1)} G_{n,m} \cos\phi_{n,m} \;. 
\end{align}
The differential mobility can be obtained either from this form
using the relation (\ref{diffmobauschi}) or from its
definition with eq.~(\ref{qlambdafc}). In any event, we arrive at the 
series expression
\begin{align}\label{mudform}
\mu_{\rm d}(\epsilon)&=\frac{q_0^2}{2\hbar}\sum_{m\;\rm even}^\infty \!\!
(-1)^{m/2-1} \!\! 
 \int_{0}^\infty\!\!\tilde{\mathcal{D}}_{0,m}(\lambda\to 0)\sum_{\{\xi_l\}}
\bigg(\prod_{l=1}^{m-1}\sin\theta_{l,m}\bigg) G_{0,m} B_{0,m}\; , \\
B_{0,m}& = \frac{\partial}{\partial\epsilon}\sin \phi_{0,m} =
\bigg(\sum_{l=1}^{m-1} g_{l,m}\rho_{l}\bigg) 
\cos\phi_{0,m}\;.   \notag
\end{align} 

As well, the mean square displacement can be
divided into the 
second moment with factorized initial state at time zero,
$\langle\hat{q}^2(\lambda)\rangle_{\text{fc}}$, and the correction
term $\hat{K}_{\text {ne}}(\lambda)$, 
\begin{align}
\label{cnelambda}
 \hat{C}_{\text {ne}}(\lambda)&=\langle\hat{q}^2(\lambda)\rangle_{\text
{fc}}+\hat{K}_{\text {ne}}(\lambda)\;,\\
\label{q2lambdafc}
\langle\hat{q}^2(\lambda)\rangle_{\text
{fc}}&=\frac{1}{\lambda^2}\sum_{m\;\rm even}^\infty
\int_{0}^\infty\!\tilde{\mathcal{D}}_{0,m}(\lambda)
\sum_{\{\xi_l\}}a_{0,m}^{(2)} G_{0,m} \cos\phi_{0,m}\;,\\
\label{knelambda}
\hat{K}_{\text{ne}}(\lambda)&=\frac{1}{\lambda} \sum_{n,m=1}^\infty
\int_{0}^\infty\!\tilde{\mathcal{D}}_{n,m}(\lambda)\int_0^\infty
\!d\tau_n\,ds_n\,e^{-\lambda s_n}
\sum_{\{\xi_l\}}a_{n,m}^{(2)} G_{n,m} \cos\phi_{n,m}\;.
\end{align}
The phase factor resulting from the $\{\eta_j\}$ summation differs from 
$a_{n,m}^{(1)}$. We have
\begin{align}
\label{aij2}
a_{n,m}^{(2)}&=-iq_0\,a_{n,m}^{(1)} \sum_{k=n+1}^{n+m-1}\cot(\theta_{k,n+m})
\;. 
\end{align}
The formally exact series expression for the noise  
follows from (\ref{sauscne}) and (\ref{cnelambda}) - (\ref{knelambda}) as 
\begin{align}
  \label{sausclambda}
S(\omega)&=2 (e^2/q_0^2) \,\mbox{Re}\,\big[\lambda^2
\hat{C}_{\text{ne}}(\lambda)\big]_{\lambda=-i\omega+0^+} \;.
\end{align}

Consider now the $\lambda$-expansions of
eqs.~(\ref{qnelambda}) - (\ref{rnelambda}) and
eqs.~(\ref{cnelambda}) - (\ref{knelambda}) about $\lambda=0$, which
determine upon Laplace transformation the behavior of the respective 
quantities at long times. 
We first observe that the phase factor $a_{n,m}^{(1)}$
vanishes if the system reaches a diagonal state
$g_{l,n+m}=0$. Therefore, all paths that contribute to the first moment
hit the diagonal of the RDM only in the initial and final
state. Thus each interval $\rho_l$ separates two
charged clusters, and for $\lambda=0$, the interaction factor 
$G_{n,m}$ at $T>0$ or the bias factor $B_{n,m}$ at $T=0$ ensures the
convergence of the respective integration. For the first
moment with factorized initial conditions, 
eq.~(\ref{qlambdafc}), we find 
\begin{align}\label{qfclowl}
\lambda^2\langle\hat q(\lambda)\rangle_{\rm fc}&=F\mu+\lambda
q_\infty+{\mathcal O}(\lambda^2)\;, & q_\infty&=\lim_{\lambda\to
0}(\partial/\partial\lambda)\big[\lambda^2\langle\hat
q(\lambda)\rangle_{\rm fc}\big]\;.
\end{align}
The series expression for $\mu$ is as in (\ref{mudform}), but
with $B_{0,m}=\sin(\phi_{0,m})/\epsilon$.
 
The series (\ref{rnelambda}) can be rearranged upon
using the relation (we put $\rho_n =\tau_n +s_n$) 
\begin{align}\label{subtrac}
\int_0^\infty d\tau_n\,ds_n\,e^{-\lambda s_n}_{}\,A (\tau_n + s_n)
&= \frac{1}{\lambda}\int_0^\infty d\rho_n\,
[\,1- e^{-\lambda\rho_n}_{}\,]\,A(\rho_n)  \; ,
\end{align}
where $A(\rho_n)$ subsumes all correlations across the intervall $\rho_n$.
It is then straightforward to see that the $\lambda$-dependence of
the expression $\lambda^2\hat R_{\rm ne}(\lambda)$ is fully cancelled
by $\lambda^2\langle\hat q(\lambda)\rangle_{\rm fc}$.
In the end, we find using eq.~(\ref{qnelambda}) for all $\lambda$
\begin{align}\label{qnelowl}
\lambda^2\langle\hat q(\lambda)\rangle_{\rm ne}&=F\mu \; .
\end{align}
This verifies that the mean position with initial value 
$\langle q(t=0)\rangle_{\rm ne} =0$
evolves in the stationary nonequilibrium state  for all times 
as $\langle q(t)\rangle_{\rm ne} = F\mu t$. 

After these preliminaries, consider now the mean square
displacement, eq.~(\ref{cnelambda}).
In the phase factor $a_{n,m}^{(2)}$, the factor $\cos(\pi K g_{l,n+m})$ 
can be allocated to any interval in the positive-time branch.
However, upon employing the relation (\ref{subtrac}), the series for
$\hat C_{\rm ne}(\lambda) = \langle\hat{q}^2(\lambda)\rangle_{\text
{fc}}+\hat{K}_{\text {ne}}(\lambda) $ can be rearranged so
that most contributions
of the sum in eq. (\ref{aij2}) cancel out. The only remaining contribution is
the one in which the cosine factor covers the first interval in the  
positive-time branch, labelled by $\rho_n$.  
Thus, this is the only interval which is
allowed to be occupied by a diagonal state, $g_{n,n+m} =0$.
Since $\rho_n$ separates two neutral $\{\xi_j\}$-clusters, the attractive 
bias and charge interaction factors are missing, and hence this interval 
gets very long as $\lambda \to 0$. It is convenient
to factorize the interaction term $G_{n,m}$ ($n$, $m$ even) into
the intra-cluster interaction factors $G_{n,0}$ and $G_{0,m}$ and
the inter-cluster interaction factor $G_{\rm int}$,
$G_{n,m}=G_{n,0}G_{0,m}G_{\rm int}$. Expanding the
inter-cluster interaction stretching over the long interval
$\rho_{n}$, we get
\begin{align}\label{gint}
G_{\rm int}&=1-\ddot{Q}'(\rho_{n})\sum_{l=1}^{n-1}g_{l,n+m}\rho_l
\sum_{k=n+1}^{n+m-1}g_{k,n+m}\rho_k\;.
\end{align}
The first term is the contribution in the absence of the cluster
interaction, and the
second term represents the dipole-dipole interaction.
 Accordingly, we get two contributions to $\hat C_{\rm
ne}(\lambda)$, called $\hat C^{(a)}_{\rm ne}(\lambda)$ and  $\hat
C^{(b)}_{\rm ne}(\lambda)$.  In the first case,
the cluster to the left of $\rho_{n}$ is just
$F\mu = \lambda^2\langle\hat q(\lambda)\rangle_{\text {ne}}$,
whereas the cluster to the right of $\rho_n$ 
represents 
$\lambda^2 \langle\hat q(\lambda)\rangle_{\text {fc}}$
[Note that each cluster is actually  an infinite series in $\Delta^2$].
 Since the two 
clusters are noninteracting, the integration over $\rho_{n}$ gives a factor 
$1/\lambda$. Thus we find
\begin{align}
\lambda^2 \hat C^{(a)}_{\rm
ne}(\lambda)&= 2\big[\, F\mu \,\big] 
\big[\lambda^2\langle\hat q(\lambda)\rangle_{\rm fc}\big]/\lambda 
= 2 F^2\mu^2/\lambda + 2F\mu q_\infty+{\mathcal
O}(\lambda)\;.\label{f1lowl} 
\end{align}
In the second form, we have used the result (\ref{qfclowl}). 

Consider next the dipole contribution $\hat C^{(b)}_{\rm
ne}(\lambda)$. Upon inserting the second  term of eq.~(\ref{gint})
into the series for $ \hat C_{\rm ne}(\lambda)$, we observe
that the two sums can be generated by suitable differentiations with
respect to the bias. Thus we find
\begin{align}\label{cnef2}
\lambda^2 \hat C^{(b)}_{\rm
ne}(\lambda)&=-\frac{q_0^2}{2\lambda^2}\sum_{n\;{\rm even}}^{\infty}\;
\sum_{m\; {\rm even}}^{\infty}\int_0^\infty d\rho_n\,e^{-\lambda\rho_n}_{} 
\ddot Q'(\rho_{n})
\int_0^\infty\tilde{\mathcal D}_{n,m}(\lambda)\, \\   \notag 
&\quad\times\sum_{\{\xi_l\}_{n}}\! \sum_{\{\xi_k\}_{m}} \! 
G_{n,0}G_{0,m} \Big(\frac{\partial}
{\partial \epsilon}\sin\phi_{n,0}\Big)  \Big(\frac{\partial}
{\partial \epsilon}\sin\phi_{0,m}\Big) \! 
\!\prod_{l=1,\,l\neq n}^{n+m-1}\!\!
\sin\theta_{l,n+m}\;. 
\end{align}
The sums $\sum_{\{\xi_l\}_n}$ and $\sum_{\{\xi_k\}_m}$ cover all
neutral charge clusters, i.e., they are constrained by 
$\sum_{l=1}^{n}\xi_l=0$ and $\sum_{l=n+1}^{n+m}\xi_l=0$. 

At $T=0$, the dipole-dipole-interaction is found from
eq.~(\ref{Qscal}) as $\ddot Q'(t)=-2K/t^2$. Substituting this form, 
the integration over $\rho_{n}$ in the limit $\lambda \to 0$ yields 
\begin{align}\label{intervalint}
\int_0^\infty d\rho_{n}\,e^{-\lambda\rho_{n}}\,\ddot
Q'(\rho_{n})\approx -2K\,\lambda \ln(\lambda/\lambda^\ast)\;.
\end{align}
Now, in leading order of $\lambda$ both the cluster on the left
and the cluster on the right of $\rho_{n}$ can be
identified with the differential mobility $\mu_{\rm d}$,
eq.~(\ref{mudform}). Thus we find
\begin{align}\label{f2lowl}
\lambda^2 \hat C^{(b)}_{\rm
ne}(\lambda)= -\,(4\hbar^2 K/q_0^2)\mu_{\rm d}^2\,\lambda
\ln(\lambda/\lambda^\ast) \,\big[1+{\mathcal O}(\lambda)] \;.
\end{align}
So far, we have considered the contributions to $\hat C_{\rm
ne}(\lambda)$ in which the period $\rho_n$ the system dwells in a diagonal
state is very long. However, there are also contributions in which the 
diagonal state is either occupied for a short period only or not
occupied at all, $g_{n,n+m} \neq 0$.
In both cases, the charges form a single 
neutral cluster with regular behavior in the limit $\lambda\to 0$.
These terms are combined in the expression
\begin{align}\label{f3lowl}
\lambda^2 \hat C^{(c)}_{\rm
ne}(\lambda)=2D'+{\mathcal O}(\lambda).
\end{align}
Putting all contributions together, the mean square
displacement takes the form
\begin{align} \label{cnelowl}
\lambda^2\hat C_{\rm
ne}(\lambda)=\frac{2}{\lambda}F^2\mu^2+(2F\mu q_\infty+2D')-\frac{4\hbar^2
K}{q_0^2}\mu_{\rm d}^2\lambda 
\ln(\lambda/\lambda^\ast) +{\mathcal O}(\lambda)\;.
\end{align}
Turning to the time regime, the expression (\ref{cnelowl}) gives 
the behavior at long times,
\begin{align}\label{cnelongt} 
C_{\rm ne}(t)&=F^2\mu^2t^2+
2Dt+ (4\hbar^2 K/q_0^2)\,\mu_{\rm d}^2 \ln(t/t^*)\;.
\end{align}
Here, all terms linear in $t$ are combined in the diffusion constant
\begin{align}
D&=\frac{1}{2}\lim_{t\to\infty}\frac{\partial}{\partial t}\big[C_{\rm
ne}(t)-\langle q(t)\rangle_{\rm ne}^2 \big]=D'+F\mu
q_\infty\;.\label{diffconst}  
\end{align}
The current correlation function at long times
is found from eqs. (\ref{current1}) and (\ref{cnelongt})  as
\begin{align}\label{currentlong}
\langle j(t)j(0)+j(0)j(t)\rangle_{\rm ne}= 2(e F\mu/q_0)^2
-\frac{2 \hbar e^2\mu_0}{\pi q_0^2} \,\frac{[\tilde\mu_{\rm d}]^2}{t^2}\;.
\end{align}
Here, we have normalized the differential mobility on the
mobility of a free Brownian particle, $\tilde \mu_{\rm d}=\mu_{\rm
d}/\mu_0$ with $\mu_0=q_0^2/2\pi\hbar K$. 

Using (\ref{sausclambda}), it is easy to commute the 
$\lambda$-expansion (\ref{cnelowl}) into the low-frequency noise spectrum.  
The $1/\lambda$-term in the expansion (\ref{cnelowl}) does not
contribute to the noise. The term of order $\lambda^0$ directly
relates the dc noise $S(\omega=0)$ to the diffusion constant,
\begin{align}
S(\omega=0)=4e^2 D/q_0^2\;.
\end{align}

In the integrable BSG model, a basis of
interacting quasiparticles which scatter off the impurity or point contact
one by one can be introduced. Upon generalizing a Landauer-type 
approach familiar from transport of free electrons to this interacting case,
it has been argued that the dc noise at $T=0$ is connected with the 
conductance by 
differentiation with respect to the voltage \cite{fendley2} or the  
tunneling amplitude \cite{solid,comment}
\begin{align}\label{dcnoise}
S(\omega=0)=eV \,\Delta\, \partial G/\partial\Delta
= [1/(K-1)] eV^2 \,\partial G/\partial V\;.
\end{align}
In order $\Delta^2$, this formula reproduces shot noise,
$S(\omega=0)=2eVG=2eI$. The higher order terms describe 
corrections due to quantum coherence for
strong tunneling. The relation (\ref{dcnoise}) can be proven
on the basis of (\ref{mudform}) - (\ref{knelambda}) and is a
consequence of (i) a generally valid detailed balance relation (DBR)
that follows from analytic properties of the pair interaction
$Q(t)$ and (ii) of a special DBR which is restricted to the scaling limit, i.e.
relies on the particular form (\ref{Qscal}) for $Q(t)$ \cite{hinweis}.
The general DBR states that at $T=0$ there is no 
real occupation of TB states higher than the initial state. The
special DBR states that only those paths for which 
all $\{\eta_j\}$ charges have equal sign, i.e., which are free of
backward hops parallel to the diagonal, contribute to the noise. 

The low-frequency correction $\Delta S(\omega)$ is captured by  the
term of order $\lambda\ln(\lambda/\lambda^\ast)$ in (\ref{cnelowl}). 
Using (\ref{sausclambda}), we obtain for the noise the nonperturbative 
universal form
\begin{align}\label{shibatb}
\Delta S(\omega)&= 2\hbar\mu_0(e^2/q_0^2)\,\tilde\mu_{\rm
d}^2\,|\omega|
=2\hbar G_0(g) \widetilde G_{\rm d}^2\,|\omega|\;.
\end{align}
In the second form, we have
used transport quantities of the related Luttinger model, i.\,e. the
dimensionless differential conductance $\widetilde G_{\rm d}=G_{\rm
d}/G_0(g)$, normalized with the static limit of the microwave conductance of 
a quantum wire or the Hall conductance, $G_0(g)=ge^2/h$. 
This result agrees with the findings of Ref.~\cite{chamon3}.
 
The algebraic $1/t^2$ decay law of the current correlation function
(\ref{currentlong}) and  the resulting $|\omega|$-singularity in the
noise spectrum (\ref{shibatb}) are signatures of the unscreened
dipole-dipole interaction. These forms
are universal since they do not depend explicitly on the parameter $K$
or $g$ and the applied bias. The parameters are only in the mobilities or
conductances which determine the prefactor of the $1/t^2$- and
$|\omega|$-law. The forms
(\ref{currentlong}) and (\ref{shibatb}) are 
analogous to the $1/t^2$ decay law of 
the position correlation function \cite{sass-w902,corr1} in the Ohmic
two-state system  and the equivalent Shiba relation in frequency space,
respectively \cite{sass-w902,corr1}. The difference, however, is in the
prefactor. For the position autocorrelation function, each of the two
sets of dipoles  
corresponds to the static susceptibility $\chi_0$, whereas in the present case
each dipole set represents the differential mobility,
which is related to the nonequilibrium susceptibility by
eq.~(\ref{diffmobauschi}). In fact, in transport problems with a stationary 
nonequilibrium state, the static susceptibility diverges, and the
differential mobility $\mu_{\rm d}(F)$ is the proper quantity here.  

At finite temperature, the dipole-dipole interaction is given by
\begin{align}\label{dipolq}
\ddot Q'(t)=-2K\,(\pi/\hbar\beta)^2/\sinh^2(\pi t/\hbar\beta)\;.
\end{align}
Performing the integration over the interval $\rho_n$
between the dipoles, eq.~(\ref{intervalint}), and switching to 
the noise, the low-frequency contribution from 
$\lambda^2\hat C^{(b)}_{\rm ne}(\lambda)$ takes the form
\begin{align}\label{shibatbT}
\Delta S(\omega,T)&=\frac{e^2}{\pi K}\big[\tilde\mu_{\rm
d}\big]^2\,\omega\coth\left(\hbar\beta\omega/2\right)
\approx\frac{2 e^2}{\pi K\hbar\beta}\big[\tilde\mu_{\rm
d}\big]^2 + \frac{e^2 \hbar\beta}{6 \pi K}\big[\tilde\mu_{\rm
d}\big]^2 \,\omega^2 +{\mathcal O}(\omega^4)\;.\notag
\end{align}
Thus, the $|\omega|$-singularity is smoothed out at finite
temperature. We get a contribution to the dc noise which is
proportional to $T$ and a correction term of order
$\omega^2$. However, there are additional contributions $\propto \omega^0$
and $\propto \omega^2$ from 
$\lambda^2\hat C^{(a)}_{\rm ne}(\lambda)$ and 
$\lambda^2\hat C^{(c)}_{\rm ne}(\lambda)$.
However, the corresponding prefactors can not be given in analytic form
for all $K$.



\begin{references}

\bibitem{wen} X. G. Wen, Phys. Rev. B {\bf 41} (1990) $12\,838$;
                Phys. Rev. B {\bf 44} (1991) 5708

\bibitem{haldane} F. D. M. Haldane, J. Phys. C {\bf 14} (1981) 3585;
     Phys. Rev. Lett. {\bf 47} (1981) 1840 

\bibitem{schulz} H. J. Schulz, in {\em Mesoscopic Quantum Physics}, Les 
Houches 1994, ed. by E. Akkermans {\it et al.},
Elsevier, Amsterdam 1995 

\bibitem{kafi} C. L. Kane and M. P. A. Fisher, Phys. Rev. B {\bf 46}
             (1992) 15\,233 

\bibitem{fendley1} P. Fendley, A. W. W. Ludwig, and H. Saleur,
           Phys. Rev. B {\bf 52} (1995)  8934 

\bibitem{kane94} C. L. Kane and M. P. A. Fisher, Phys. Rev. Lett.
                {\bf 72} (1994)  724 

\bibitem{chamon1} C. de C. Chamon, D. E. Freed, and X. G. Wen,
         Phys. Rev. B {\bf 51} (1995)  2363 

\bibitem{glattli} L. Saminadayar, D. C. Glattli, Y. Lin, and
B. Etienne, Phys. Rev. Lett. {\bf 79} (1997)  2526 

\bibitem{weizmann} R. de Picotto {\it et al.}, 
Nature (London) {\bf 389} (1997) 162

\bibitem{fendley2} P. Fendley, A. W. W. Ludwig, and H. Saleur,
          Phys. Rev. Lett. {\bf 75} (1995)  2196 

\bibitem{solid} U. Weiss,   Solid State Comm. {\bf 100} (1996)  281 

\bibitem{saleur} F. Lesage and H. Saleur, Nucl. Phys. B {\bf 490} (1997)  543

\bibitem{chamon3} C. Chamon and D. E. Freed, Phys. Rev. B {\bf 60}
(1999)  1842

\bibitem{schmid} A. Schmid, Phys. Rev. Lett. {\bf 51}  (1983) 1506

\bibitem{wbuch} U. Weiss, {\it Quantum Dissipative Systems},
      Series in Modern Condensed Matter Physics, Vol. 10,  second edition,
      World Scientific, Singapore  1999

\bibitem{schoen} G. Sch\"on and A. D. Zaikin, Phys. Reports {\bf 198} (1990) 
237 

\bibitem{ing} G.-L. Ingold and Yu. V.  Nazarov, in  {\it Single Charge 
Tunneling}  ed. by H. Grabert and M. H. Devoret, Plenum Press, New
York 1992 

\bibitem{sass-w902} M. Sassetti and U. Weiss, Phys. Rev. Lett. {\bf 65} 
(1990) 2262
                    
\bibitem{corr1} G. Lang, E. Paladino, and U. Weiss,
Europhys. Lett. {\bf 43} (1998)  117 

\bibitem{comment} In the scaling limit, we have $G= G(\Delta^2 V^{2K-2})$
and thus $V\partial G/\partial V
= [\,K-1\,]\Delta \partial G/\partial \Delta$

\bibitem{hinweis} Details will be published elsewhere

\end{references}
\end{document}